# Conditional statistical properties of the complex systems having long-range interactions

Zhifu Huang[1,*], Congjie Ou[1], Bihong Lin[1], Guozhen Su[2], Jincan Chen[2]

[1]College of Information Science and Engineering, Huaqiao University, Xiamen 361021, People's Republic of China

[2]Department of Physics, Xiamen University, Xiamen 361005, People's Republic of China

A new concept of the available force is proposed to investigate the performance of the complex systems having long-range interactions. Since the covariance of average velocity in double time interval and available force equals zero, it is possible to calculate the conditional probability distribution function (CPDF) within the systems. It is found that the asymmetric CPDF of the velocity between two adjacent time intervals can be derived from the symmetrical CPDF between the available force and the double time interval velocity. Two typical currency exchange databases, i.e., EUR/USD and GBP/USD, which collect the minutely opening exchange prices from 1 January 1999 to 31 December 2011, are adopted as examples. It is found that the analytical CPDF needs only six parameters for an arbitrary system. By calculating the CPDF in the currency exchange databases, it is shown that the results are well fitted by our analytical expression. The analytical CPDF can be also used to calculate the conditional expectation and the conditional variance of velocity. Interestingly, the two databases show that the conditional expectation of the velocity between two adjacent time intervals is not monotonic, while the conditional variance tends to monotonic. All of these results are well described by our theory. It is worthwhile to note that the analytical CPDF is a general expression. It is valid not only for current exchange systems but also for any complex systems having long-range interactions and/or long-duration memory.

**Key words**: Complex system; Long-range interaction; Conditional probability distribution function

**PACS numbers**: 05.65.+b; 05.45.Ac; 02.50.Cw

---

*Email: zfhuang@hqu.edu.cn

The charm of science is forecasting future outcomes on current information. As there exists the complicated dynamics in complex systems, it is hand to obtain deterministic prediction. However, we can also explore the prediction of statistical properties in complex systems. First of all, in order to forecast the statistical properties in complex systems, we must analyze the dynamics from real complex systems. For the sake of simplicity, we focus on the one-dimensional case below. In real systems, the time interval of any process is finite. When the position is ensured, the displacement during the time interval $\Delta t$ can be expressed as

$$x(t,\Delta t) = s(t+\Delta t) - s(t)\,, \tag{1}$$

where $s(t)$ is the position at time $t$ and $\Delta t$ is the time interval. At the same time, one can define the corresponding velocity as

$$v(t,\Delta t) = \frac{x(t,\Delta t)}{\Delta t} = \frac{s(t+\Delta t) - s(t)}{\Delta t}\,. \tag{2}$$

It is important to note that the displacement and velocity mentioned here is not the instantaneous value. The displacement and velocity depend not only on time but also on time interval. Therefore, we can analyze the displacement and velocity in different time intervals.

A typical time interval dependent series is the particle's displacement during anomalous diffusion. Anomalous diffusion is a phenomenon encountered in almost every branch of science. Many physical, biological, and finical systems [1]-[4] that contain fractal and self-similar structures, long-range interaction, and/or long-duration memory, have anomalous diffusions. These anomalous diffusions can be characterized by one-dimensional mean-square displacement as

$$\sigma^2_{x(t,\Delta t)} = \left\langle x^2(t,\Delta t) \right\rangle \propto \Delta t^{\alpha}\,, \tag{3}$$

where $\sigma^2_{x(t,\Delta t)}$ is called the mean-square displacement or variance of displacement, and $\alpha$ is the diffusion coefficient. The cases of $\alpha < 1$ and $\alpha > 1$ correspond to the subdiffusion and superdiffusion, respectively, while $\alpha = 1$ corresponds to the normal diffusion or Brownian motion. Using equations (2) and (3), one can obtain the mean-square velocity during time interval $\Delta t$ as

$$\sigma^2_{v(t,\Delta t)} = \left\langle v^2(t,\Delta t) \right\rangle \propto \Delta t^{\alpha-2}, \tag{4}$$

where $\sigma^2_{v(t,\Delta t)}$ is called the mean-square velocity or variance of velocity. From equations (3) and (4), it can be seen that when $0 < \alpha < 2$, the mean-square displacement is convergent but the mean-square velocity is divergent at $\Delta t \to 0$. Since the mean-square velocity is divergent at $\Delta t \to 0$, the instantaneous velocity can not be strictly measured. Therefore, a moving object may need an infinite force to change its speed or direction discontinuously. In this situation, neither the instantaneous force nor the mean force can be defined. According to Newton's law, we can define an available force as

$$F(t,\Delta t) = m\frac{v(t+\Delta t,\Delta t) - v(t,\Delta t)}{\Delta t}, \tag{5}$$

where *m* is the mass of the moving object and we set *m*=1 for the sake of convenience. It is worthwhile to note that in different time intervals, the available force has different values. Presently, we proposed a new method of analyzing the available force to describe the long-range interaction complex systems [5]. It is found that the long-range interaction, anomalous diffusion, and *q*-Gaussian shape distribution can be well described by the interaction parameters in different cases, where the velocity of each case is divergent and integrable everywhere. From equation (2), one can easily obtain the expression of velocity at time *t* in double time interval as

$$v(t,2\Delta t)=\frac{v(t,\Delta t)+v(t+\Delta t,\Delta t)}{2}. \tag{6}$$

From equations (5) and (6), one can find that if $v(t,\Delta t)$=$y$ and $v(t+\Delta t,\Delta t)$=z are defined, $v(t,2\Delta t)$=$(y+z)/2$ and $F(t,\Delta t)=(z-y)/\Delta t$ are yielded. As a result, the joint probability distribution functions (JPDF) of events $v(t,\Delta t)$=$y$ and $v(t+\Delta t,\Delta t)$=z are equal to the JPDF of events $v(t,2\Delta t)$=$(y+z)/2$ and $F(t,\Delta t)=(z-y)/\Delta t$. Then, we can obtain

$$p_{v(t,\Delta t),v(t+\Delta t,\Delta t)}(y,z)=p_{v(t,2\Delta t),F(t,\Delta t)}(\frac{y+z}{2},\frac{z-y}{\Delta t}), \tag{7}$$

where $p_{v(t,\Delta t),v(t+\Delta t,\Delta t)}(y,z)$ is the JPDF of events $v(t,\Delta t)$=$y$ and $v(t+\Delta t,\Delta t)$=z and $p_{v(t,2\Delta t),F(t,\Delta t)}[(y+z)/2,(z-y)/\Delta t]$ is the JPDF of events $v(t,2\Delta t)$=$(y+z)/2$ and $F(t,\Delta t)=(z-y)/\Delta t$. For the sake of convenience, $A(t+m\Delta t,\mathrm{n}\Delta t)$ is simplified as $\mathrm{A}_{m,n}$. For example, $v(t,\Delta t)$, $v(t+\Delta t,\Delta t)$, $v(t,2\Delta t)$, and $F(t,\Delta t)$ may be, respectively, simplified as $v_{0,1}$, $v_{1,1}$, $v_{0,2}$, and $F_{0,1}$.

It is well known that the JPDF of two events can be calculated as the probability distribution function (PDF) of the first event multiplies the conditional probability distribution function (CPDF) of the next event in the condition of the first event happened. Then, one can obtain the expression as

$$p_{v_{0,1},v_{1,1}}(y,z)=p_{v_{0,1}}(y)\,p_{v_{1,1}|v_{0,1}}(z\,|\,y) \tag{8}$$

and

$$p_{v_{0,2},F_{0,1}}(y,z)=p_{v_{0,2}}(y)\,p_{F_{0,1}|v_{0,2}}(z\,|\,y), \tag{9}$$

where $p_{v_{0,1}}(y)$ and $p_{v_{0,2}}(y)$ are the PDFs of events $v(t,\Delta t)$=$y$ and $v(t,2\Delta t)$=$y$, respectively, while $p_{v_{1,1}|v_{0,1}}(z\,|\,y)$ is the CPDF of event $v(t+\Delta t,\Delta t)$=z in the condition of event $v(t,\Delta t)$=$y$, and $p_{F_{0,1}|v_{0,2}}(z\,|\,y)$ is the CPDF of event $F(t,\Delta t)=z$ in the condition of event $v(t,2\Delta t)$=$y$.

According to the Bayesian method, the prior CPDF or posterior CPDF can be used to calculate the other CPDF. Substituting equations (8) and (9) into equation (7), it is very natural to obtain the following relation

$$p_{v_{1,1}|v_{0,1}}(z\,|\,y)\text{=}p_{v_{0,2}}(\frac{y+z}{2})\,p_{F_{0,1}|v_{0,2}}(\frac{z-y}{\Delta t}\,|\,\frac{y+z}{2})\,/\,p_{v_{0,1}}(y)\,. \tag{10}$$

From equation (10), we can calculate CPDF $p_{v_{1,1}|v_{0,1}}(z\,|\,y)$ from CPDF $p_{F_{0,1}|v_{0,2}}(z\,|\,y)$. On the other hand, one can calculate the covariance of $v(t,2\Delta t)$ and $F(t,\Delta t)$ from equations (5) and (6) as

$$\left\langle v(t,2\Delta t)F(t,\Delta t)\right\rangle=\frac{1}{2\Delta t}[\left\langle v^2(t+\Delta t,\Delta t)\right\rangle\text{-}\left\langle v^2(t,\Delta t)\right\rangle]\,. \tag{11}$$

Fortunately, equation (11) is equal to zero because the mean-square velocity (4) does not depend on time *t*. Thus, we can find that the incidence between $v(t,2\Delta t)$ and $F(t,\Delta t)$ is neither positive nor negative. Therefore, we can assume that the CPDF $p_{F_{0,1}|v_{0,2}}(z\,|\,y)$ is a symmetrical function. After that, we may only analyze the conditional variance and tail of CPDF $p_{F_{0,1}|v_{0,2}}(z\,|\,y)$ to obtain its expression. Finally, we can obtain CPDF $p_{v_{1,1}|v_{0,1}}(z\,|\,y)$ from CPDF $p_{F_{0,1}|v_{0,2}}(z\,|\,y)$. From equation (4), one can also find that the covariance of $v(t,\Delta t)$ and $v(t+\Delta t,\Delta t)$ is not equal to zero when the diffusion coefficient $\alpha$ is not equal to 1. Therefore, the CPDF $p_{v_{1,1}|v_{0,1}}(z\,|\,y)$ is an asymmetric function, and the form of CPDF $p_{v_{1,1}|v_{0,1}}(z\,|\,y)$ may be more complex than that of CPDF $p_{F_{0,1}|v_{0,2}}(z\,|\,y)$. As a result, it is possible and beneficial to obtain asymmetric CPDF $p_{v_{1,1}|v_{0,1}}(z\,|\,y)$ from symmetrical CPDF $p_{F_{0,1}|v_{0,2}}(z\,|\,y)$ in long-range interaction complex systems.

The PDF of velocity in long-range interaction complex systems usually does not agree with the Gaussian distribution coming from independent or nearly independent contributions, but may take the forms of non-Gaussian distribution, e.g., Levy stable forms [1]-[4], stretched Gaussian shape [4], [6], or the form of *q*-Gaussian, which is given by

$$p(y) = C[1-(1-q)\beta y^2]^{1/(1-q)}, \qquad (12)$$

where $\beta$ is a parameter characterizing the width of the distribution and $q$ is the nonextensivity index [7], [8], while C is the normalized parameter. In equation (12), $q \neq 1$ indicates a departure from the Gaussian shape while $q \to 1$ limit yields the normal Gaussian distribution. When $q > 1$ and $y^2 >> 1/[(q-1)b]$, equation (12) may be simplified into the power-law shape, i.e.,

$$p(y) \propto y^{\gamma}, \qquad (13)$$

where $\gamma = 2/(1-q)$. It means that the *q*-Gaussian form of PDF may be used to analyze the asymptotic of Lévy stable forms [9]. On the other hand, the normalized parameter C depends on $\beta$ and $q$ and can be expressed as

$$C = \frac{\sqrt{\beta(\mathrm{q}-1)}\,\Gamma[1/(q-1)]}{\sqrt{\pi}\,\Gamma[(3-q)/(2q-2)]}, \qquad (14)$$

where $\Gamma(...)$ is the gamma function, while $q$ must be in the range of $1 < q < 3$ to validate the gamma function. In addition, the variance of equation (12) also depends on $\beta$ and $q$, which is given by

$$\sigma^2 = \frac{\Gamma[(5-3q)/(2q-2)]}{2\beta(\mathrm{q}-1)\,\Gamma[(3-q)/(2q-2)]}, \qquad (15)$$

where $q$ must be in the range of $1 < q < 1.6$ to keep the gamma function valid. From equations (14) and (15), it can be seen that both $C$ and $\sigma$ are determinate if $\beta$ and $q$ are given. Mathematically speaking, there are only two independent parameters among $\beta$, $q$, $C$, and $\sigma$, so that we can choose arbitrary two among them to describe the PDF. For example, we can choose $q_{v_{0,1}}$ and $\sigma_{v_{0,1}}$ to describe the PDF of velocity in one time interval.

The statistical properties of currency market fluctuations are important for modeling and understanding complex market dynamics. We analyze two typical currency exchange databases of Euro vs U.S. Dollar (EUR/USD) and Great Britain Sterling Pound vs U.S. Dollar (GBP/USD), which can be found at the following websites: http://finance.yahoo.com and http://www.metaquotes.net. We collect the minutely opening exchange prices from 1 January 1999 to 31 December 2011, so the time interval is 1 minute for the data sequences. For the sake of convenience, we can normalize $\Delta t$ as 1 in our work and also define a basic quantity [10]: the position $s(t)$ is the logarithmic of the exchange price $s(t)=\log[price(t)]$. After that, the corresponding displacement, velocity and available force can be obtained. In order to simplify our database, the obtained velocity is normalized to the standard deviation $\sigma_{v_{0,1}}$, while $\sigma_{v_{0,1}}$ equals $2.194\times10^{-4}$ and $1.932\times10^{-4}$ in EUR/USD and GBP/USD, respectively. After the database has been simplified, one can obtain that $\sigma_{v_{0,2}}$ equals 0.654 and 0.663 in EUR/USD and GBP/USD, respectively. Figures 1a and 1c show that the $q$-Gaussian distribution can be well approximated by the PDF of velocity and available force for the data, while different cases have different values of $q$.

If there does not exist long-range interaction in the sequence, the CPDF $p_{F_{0,1}|v_{0,2}}(\mathrm{z|y})$ is independent of $v(t,2\Delta t)=\mathrm{y}$. Otherwise, the dependence of CPDF $p_{F_{0,1}|v_{0,2}}(\mathrm{z|y})$ on $v(t,2\Delta t)=\mathrm{y}$ will reveal the long-range interaction within the systems. Since the covariance of $v(t,2\Delta t)$ and $F(t,\Delta t)$ is equal to zero, we can assume that the corresponding CPDF $p_{F_{0,1}|v_{0,2}}(\mathrm{z|y})$ also satisfies the $q$-Gaussian shape and it can

be written as

$$p_{F_{0,1}|v_{0,2}}(\mathrm{z}|\mathrm{y}) = \mathrm{C}_{F_{0,1}|v_{0,2}}(\mathrm{y})\{1\text{-}[1\text{-}\mathrm{q}_{F_{0,1}|v_{0,2}}(y)]\beta_{F_{0,1}|v_{0,2}}(y)z^2\}^{1/[1\text{-}\mathrm{q}_{F_{0,1}|v_{0,2}}(y)]}, \quad (16)$$

where $\mathrm{C}_{F_{0,1}|v_{0,2}}(\mathrm{y})$, $\mathrm{q}_{F_{0,1}|v_{0,2}}(y)$ and $\beta_{F_{0,1}|v_{0,2}}(y)$ are conditional parameters which depend on $v(t,2\Delta t)=\mathrm{y}$. It should be mentioned that there are only two independent parameters in the system, so we can choose $\sigma^2_{F_{0,1}|v_{0,2}}(y)$ and $q_{F_{0,1}|v_{0,2}}(y)$ to analyze the CPDF $p_{F_{0,1}|v_{0,2}}(\mathrm{z}|\mathrm{y})$, where $\sigma^2_{F_{0,1}|v_{0,2}}(y)$ is the conditional variance that depends on $v(t,2\Delta t)=\mathrm{y}$ and $q_{F_{0,1}|v_{0,2}}(y)$ describes the fat-tail of the distribution. It can be seen from figures 1b and 1d that the *q*-Gaussian distribution can be well approximated by a symmetrical CPDF $p_{F_{0,1}|v_{0,2}}(\mathrm{z}|\mathrm{y})$. For different values of y, the corresponding *q* values are also different. Thus, it is possible to describe the CPDF $p_{F_{0,1}|v_{0,2}}(\mathrm{z}|\mathrm{y})$ in long-range interaction complex systems by *q*-Gaussian distributions.

Because complex systems sometimes exhibit long-range interaction, the relationship of CPDF between the available force and the double time interval velocity may not be easily obtained. However, an intriguing aspect of the complex systems is to exhibit self-similar structures [11] characterized by scale invariance and plays a central role in a large number of physics phenomena [12]. As the time interval of velocity $v(t,2\Delta t)$ includes the time interval of available force $F(t,\Delta t)$, we may suppose that the velocity $v(t,2\Delta t)$ and the available force $F(t,\Delta t)$ in the same quasi-system may include self-similar structures. As a result, the conditional variance between the available force and the double time interval velocity can be written as a linear function of $y^2$, i.e.,

$$\sigma^2_{F_{0,1}|v_{0,2}}(y) = r_{\sigma 0} + r_{\sigma 1}y^2, \quad (17)$$

where $r_{\sigma 0}$ and $r_{\sigma 1}$ are two parameters. It can be seen from figures 2a and 2c that equation (17) can be well fitted by the real data.

In addition, we can analyze the conditional $q$ of available force versus velocity in double time interval. Since the covariance of $v(t, 2\Delta t)$ and $F(t, \Delta t)$ equals zero, the conditional $q$ between the available force and the double time interval velocity may depend on the square of velocity in double time interval. Furthermore, it should be mentioned that when the velocity is increased, the corresponding probability of available force will decrease and the corresponding conditional available force will tend to being independent from each other. Thus, when the velocity in double time interval is large enough, the conditional $q$ will tend to 1. For the sake of convenience, if the conditional $q$ is less than 1.01, we set it as 1.01. Therefore, the range of velocity in double time interval can not be too large and we fit the conditional $q$ values by the following expression,

$$q_{F_{0,1}|v_{0,2}}(y) = r_{q0} + r_{q1} y^2, \tag{18}$$

where $r_{q0}$ and $r_{q1}$ are two parameters of the function $q_{F_{0,1}|v_{0,2}}(y)$. It is shown in figures 2b and 2d that equation (18) is well fitted by the real data. Thus, we can find that the $q$-Gaussian distribution, conditional variance and conditional $q$ values are enough to describe the CPDF $p_{F_{0,1}|v_{0,2}}(\mathrm{z|y})$. Moreover, substituting equations (14), (15), (17), and (18) into equation (16), the CPDF $p_{F_{0,1}|v_{0,2}}(\mathrm{z|y})$ can be explicitly obtained, and equation (10) can be used to obtain CPDF $p_{v_{1,1}|v_{0,1}}(z \mid y)$. It is important to note that the CPDF $p_{v_{1,1}|v_{0,1}}(z \mid y)$ in all different cases can be explicitly expressed with only six parameters, $q_{v_{0,2}}$, $\sigma_{v_{0,2}}$, $r_{\sigma 0}$, $r_{\sigma 1}$, $r_{q0}$ and $r_{q1}$, which can be obtained from the data fitting.

For example, we can calculate the CPDF $p_{v_{1,1}|v_{0,1}}(z|y)$ in the condition of $v(t,\Delta t)=0$ and $v(t+\Delta t,\Delta t)=z$, which implies that the object is static in the first time interval and is moving with velocity $z$ in the next time interval. From equations (5) and (6), one can obtain the corresponding values of $F(t,\Delta t)=z/\Delta t$ and $v(t,2\Delta t)=\mathrm{z}/2$, so that equation (16) can be simplified as

$$p_{F_{0,1}|v_{0,2}}(\frac{\mathrm{z}}{\Delta t}|\frac{z}{2}) \\ =\mathrm{C}_{F_{0,1}|v_{0,2}}(\frac{z}{2})\{1\text{-}[1\text{-}\mathrm{q}_{F_{0,1}|v_{0,2}}(\frac{z}{2})]\beta_{F_{0,1}|v_{0,2}}(\frac{z}{2})(\frac{z}{\Delta t})^2\}^{1/[1\text{-}\mathrm{q}_{F_{0,1}|v_{0,2}}(\frac{z}{2})]}. \tag{19}$$

From equation (19), the extreme situation $v(t+\Delta t,\Delta t)\gg 0$ can be analyzed. Because the situation $v(t+\Delta t,\Delta t)\gg 0$ implies $z\gg 0$ in equation (19), one can obtain $\beta_{F_{0,1}|v_{0,2}}(\mathrm{z})\propto \mathrm{z}^{-2}$ from equations (15) and (17). By substituting $\beta_{F_{0,1}|v_{0,2}}(\mathrm{z})\propto \mathrm{z}^{-2}$ and equation (14) into equation (19), the CPDF $p_{F_{0,1}|v_{0,2}}(\mathrm{z}/\Delta \mathrm{t}|z/2)$ can be simplified as

$$p_{F_{0,1}|v_{0,2}}(\frac{\mathrm{z}}{\Delta t}|\frac{z}{2})\propto z^{-1}. \tag{20}$$

Comparing equations (10) and (12) with equation (20), we can obtain the CPDF $p_{v_{1,1}|v_{0,1}}(\mathrm{z}|0)$ as

$$p_{v_{1,1}|v_{0,1}}(\mathrm{z}|0)\propto (z^2)^{1/[1-(3q_{v_{0,2}}-1)/(q_{v_{0,2}}+1)]}. \tag{21}$$

From equation (21), one can obtain

$$\mathrm{q}_{v_{1,1}|v_{0,1}}(0)=(3q_{v_{0,2}}-1)/(q_{v_{0,2}}+1), \tag{22}$$

where $\mathrm{q}_{v_{1,1}|v_{0,1}}(0)$ is the conditional $q$ of velocity in the next time interval and the condition is that in the first time interval, the velocity equals zero. It can be seen that $\mathrm{q}_{v_{1,1}|v_{0,1}}(0)$ depends on $\mathrm{q}_{v_{0,2}}$ that represents the velocity distribution in double time interval. From equation (22), one can find that the inequality $q_{v_{1,1}|v_{0,1}}(0) < q_{v_{0,2}}$ will be established in the range of $1 < q < 1.6$. On the other hand, from equation (12) we can see that *q* represents the "tail" of the distribution or the extreme probability of system. The extreme probability will decrease when *q* is decreased. Therefore, it is important to note that the conditional extreme probability of velocity in the next time interval in the condition $v(t,\Delta t) = 0$ is always less than the extreme probability of velocity without conditions, which requires $q_{v_{1,1}|v_{0,1}}(0) < q_{v_{0,2}}$. Besides, it can be seen from figures 1a and 1c that the *q* value of the velocity distribution tends to fixedness if the time interval is doubled, i.e. $q_{v_{0,2}} \approx q_{v_{0,1}}$. The difference between the PDF $p_{v_{0,1}}(y)$ and the CPDF $p_{v_{1,1}|v_{0,1}}(z|0)$ in figures 3a and 3b shows clearly that the extreme probabilities of $p_{v_{1,1}|v_{0,1}}(z|0)$ are less than those of $p_{v_{0,1}}(y)$, and the conditional $q_{v_{1,1}|v_{0,1}}(0)$ equals 1.34 which satisfies equation (22).

Moreover, using equations (10) and (12)-(18), we can obtain the analytical CPDF $p_{v_{1,1}|v_{0,1}}(z|y)$ and calculate its values in all different cases. It is shown clearly in figures 3b and 3d that the theoretical curves are well approximated by the data in different cases, and CPDF $p_{v_{1,1}|v_{0,1}}(z|y)$ is an asymmetric function when $y \neq 0$. The above results are obtained from the database of currency exchange rates but the framework is beyond the economic phenomena. It is important to note that the analytical CPDF is a general expression, which may allow us to obtain the CPDF in other long-range interaction complex systems with the same method.

The analytical expression of the conditional expectation of velocity between two adjacent time intervals can be expressed as

$$M_{v_{1,1}|v_{0,1}}(\mathrm{y}) = \int_{W_z|y} z p_{v_{1,1}|v_{0,1}}(z \mid y) dz \,, \tag{23}$$

where $W_z|\mathrm{y}$ means all situations in the condition $v(t, \Delta t) = y$. Since $M_{v_{1,1}|v_{0,1}}(\mathrm{y})$ depends on $v(t, \Delta t) = y$, we can numerically calculate $M_{v_{1,1}|v_{0,1}}(\mathrm{y})$ by using the obtained CPDF $p_{v_{1,1}|\mathrm{v}_{0,1}}(z \mid y)$. It can be seen from figures 4a and 4c that the theoretical curves are well approximated by the data in different conditional expectation cases. As a consequence of the long-range interaction, the relationship of conditional expectation is not monotonic and the mean velocity of the next time interval depends closely on the velocity in the previous time interval.

By the same trick, we can also numerically calculate the conditional variance of velocity between two adjacent time intervals by the following analytical formula

$$\sigma^2_{v_{1,1}|v_{0,1}}(\mathrm{y}) = \int_{W_z|y} z^2 p_{v_{1,1}|v_{0,1}}(z \mid y) dz \,. \tag{24}$$

It shows clearly in figures 4b and 4d that the theoretical curves are well approximated by the data in different conditional variance cases. These figures clearly display the effect of the long-range interaction. That is, small conditional variance is more likely to be followed by small velocity and large variance is followed by large velocity. In comparison with the previous work of Engle [13] and Bollerslev [14], the theory of conditional expectation and conditional variance presented here is based on the CPDF that can be analytically expressed. It is possible to calculate other conditional statistical properties based on the analytical CPDF. This method may be helpful to more accurate predictions of statistical properties in long-range interaction complex systems. Furthermore, the method presented here may be applicable to the analysis of a wide range of phenomena, including natural, artificial, financial and social complex systems.

To sum up, we have proposed a new method to obtain the conditional statistical properties in long-range interaction complex systems. By analyzing two typical currency exchange database of EUR/USD and GBP/USD, we can describe the CPDF $p_{v_{1,1}|v_{0,1}}(z|y)$ only with six parameters $q_{v_{0,2}}$, $\sigma_{v_{0,2}}$, $r_{\sigma 0}$, $r_{\sigma 1}$, $r_{q0}$ and $r_{q1}$ in all different cases. Moreover, we numerically calculate the conditional expectation and the conditional variance by using the obtained CPDF $p_{v_{1,1}|v_{0,1}}(z|y)$. It is found that the relationship of conditional expectation versus velocity in the first time interval is not monotonic, while the variation of conditional variance is monotonic. It is important to note that the theory of conditional expectation and conditional variance presented here is based on the CPDF obtained before. It is possible to calculate other conditional statistical properties base on the obtained CPDF. Therefore, the present results provide some helpful elements for the further understanding of the occurrence of conditional statistical properties in many natural, artificial, financial and social complex systems. It may be expected that the further research in this direction will open new perspectives and shed new light on the prediction of statistical properties in long-range interaction complex systems.

**Acknowledgments**

Project supported by the National Natural Science Foundation (No. 11247265, 11005041), the Fujian Provincial Natural Science Foundation (No. 2011J01012), the program for prominent young Talents in Fujian Province University (JA12001), and the Science Research Fund of Huaqiao University (No.11BS207), People's Republic of China.

Figure captions:

Fig.1. The PDF of velocity and available force and the CPDF between the available force and the double time interval velocity.

Fig.2. The conditional variance between the available force and the double time interval velocity for the parameters $r_{\sigma 0}$ and $r_{\sigma 1}$ and the conditional q between the available force and the double time interval velocity for the parameters $r_{q0}$ and $r_{q1}$.

Fig.3. The CPDF of the velocity between two adjacent time intervals. Dots and curves correspond to the cases of data and theory, respectively.

Fig.4. The conditional expectation of the velocity between two adjacent time intervals and the conditional variance of the velocity between two adjacent time intervals. Square dots and solid curves correspond to the cases of data and theory, respectively.

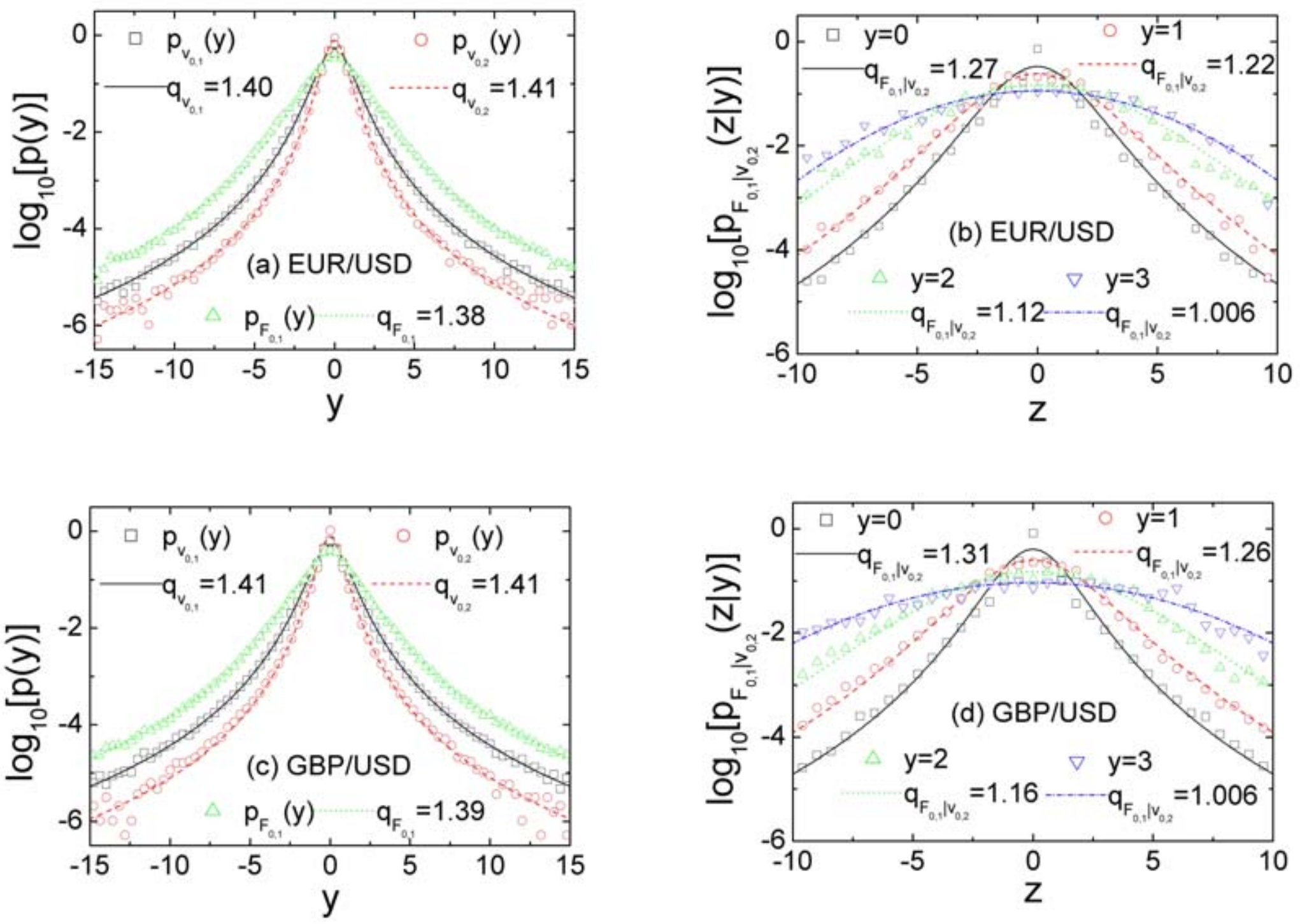

Fig.1.

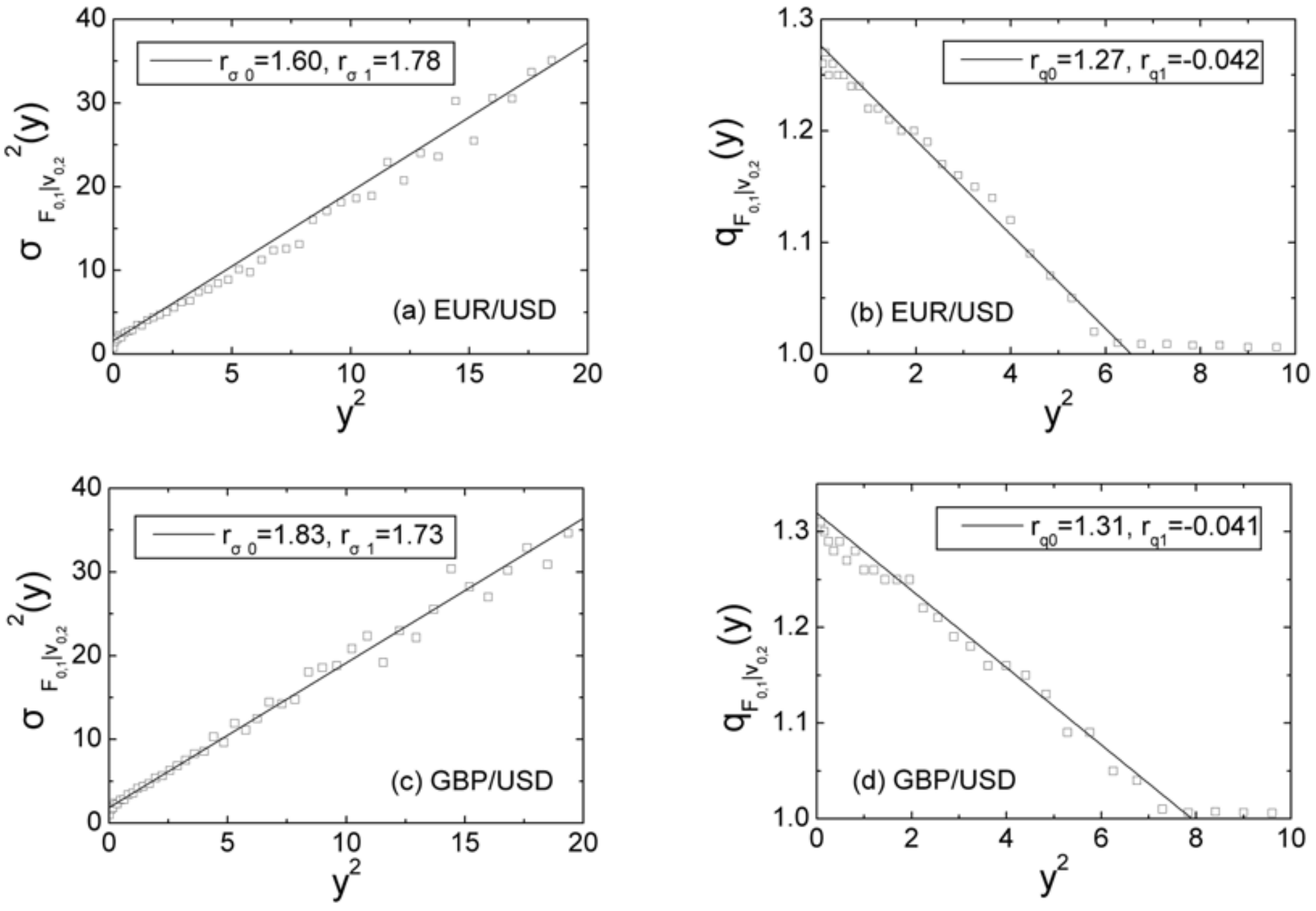

Fig.2.

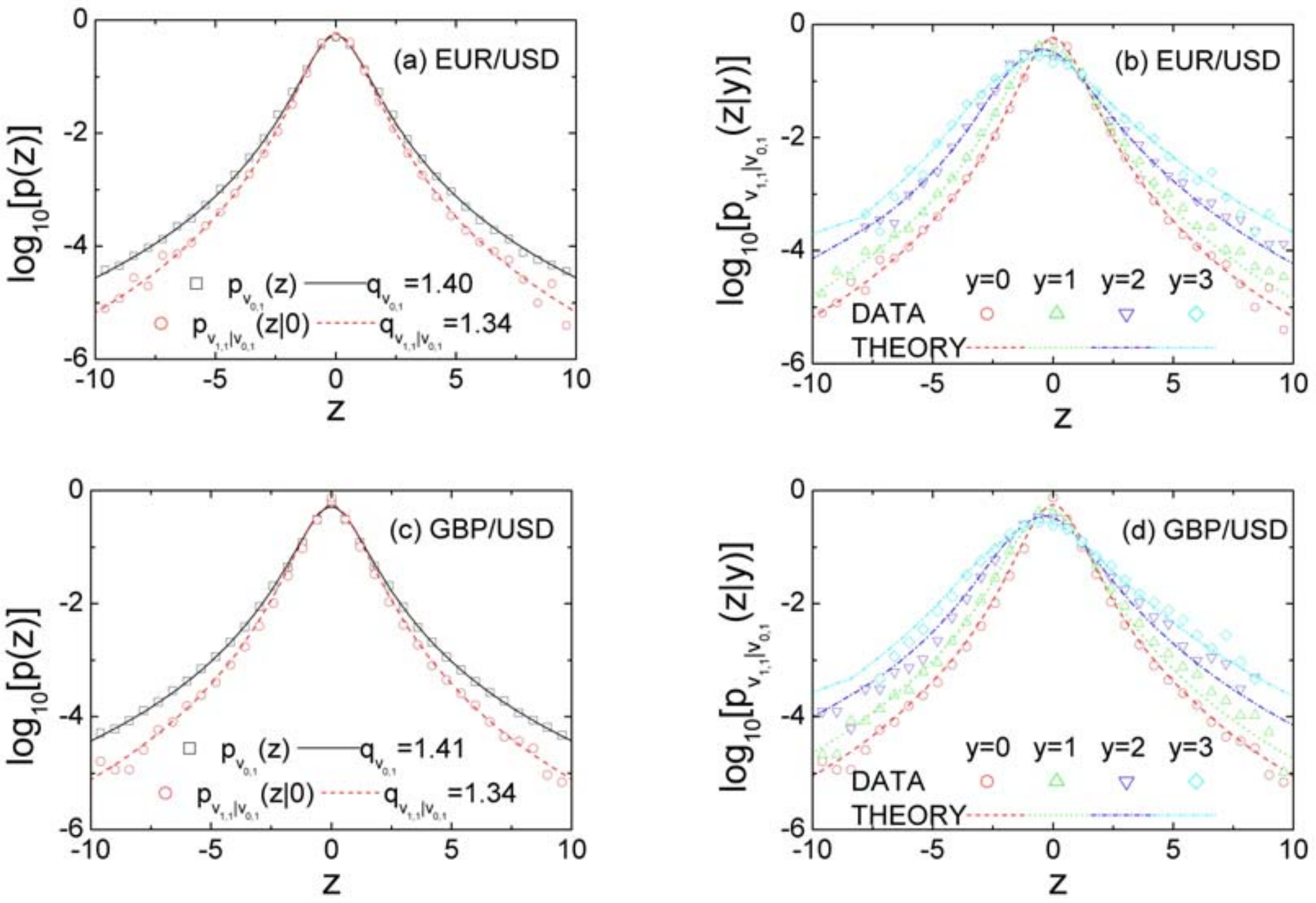

Fig.3.

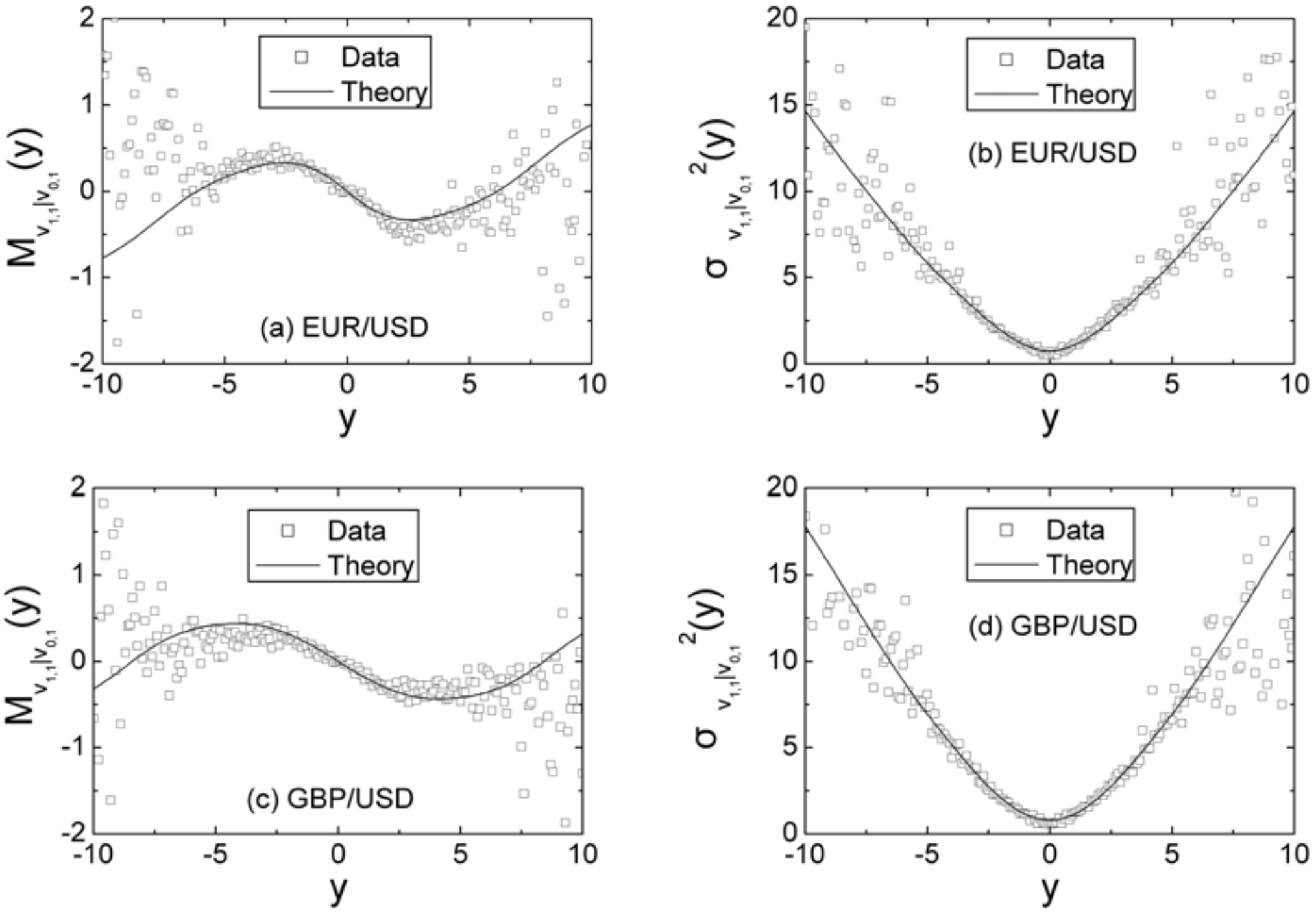

Fig.4.